\documentclass[doublecol]{epl2}
\usepackage{color,calc,graphicx,amsfonts,amscd,epsfig,psfrag,amsmath,amssymb}
\usepackage[sections,floats,textmath,displaymath]{}
\usepackage[percent]{overpic}

\renewcommand{\b}{\beta}

\newcommand{\trel}{T_{\rm rel}}

\newcommand{\D}{\Delta}

\renewcommand{\b}{\beta}
\renewcommand{\l}{\lambda}
\renewcommand{\L}{\Lambda}

\renewcommand{\l}{\lambda}
\renewcommand{\a}{\alpha}

\renewcommand{\t}{\tau}

\renewcommand{\O}{\Omega}

\newtheorem{maintheorem}{Theorem}

\newcommand{\cC}{\ensuremath{\mathcal C}}

\newcommand{\cE}{\ensuremath{\mathcal E}}

\newcommand{\cK}{\ensuremath{\mathcal K}}

\newcommand{\cT}{\ensuremath{\mathcal T}}

\newcommand{\bbN}{{\ensuremath{\mathbb N}} }

\newcommand{\bbP}{{\ensuremath{\mathbb P}} }

\newcommand{\bbZ}{{\ensuremath{\mathbb Z}} }

\let\a=\alpha \let\b=\beta     
  \let\h=\eta    \let\k=\kappa  \let\l=\lambda
          \let\p=\pi  
  \let\s=\sigma \let\t=\tau   
  
\let\D=\Delta     \let\L=\Lambda 
\let\O=\Omega      

\usepackage[normalem]{ulem}

\title{The influence of dimension on the relaxation process of East-like models}
\author{P. Chleboun\inst{1} \and A. Faggionato\inst{2} \and F. Martinelli\inst{3}}
\shortauthor{P. Chleboun \etal}

\institute{                    
  \inst{1} Mathematics Institute, University of Warwick - Coventry CV4 7AL, UK. e--mail:
paul@chleboun.co.uk\\
  \inst{2} Department of Mathematics, La
  Sapienza University - P.le Aldo Moro  2, 00185  Rome, Italy. e--mail:
  faggiona@mat.uniroma1.it\\
\inst{3}Department of Mathematics and Physics,
   Roma Tre University - Largo S.L. Murialdo 1, 00146 Rome, Italy. e--mail:
martin@mat.uniroma3.it 
}

\abstract{We consider the relaxation process and the
  out-of-equilibrium dynamics of natural generalizations to arbitrary
  dimensions of the well known one dimensional East process. 
These facilitated models
are supposed to catch some of the main features of the complex
dynamics of fragile glasses.
We focus on the low temperature regime (small density $c\approx
e^{-\b}$  of the facilitating sites).
In the literature the relaxation process  for the above models has been
assumed to be \emph{quasi-one dimensional} and,  in particular, their equilibration time has been
computed using the relaxation time of the East model ($d=1$) on the equilibrium length
scale $L_c=(1/c)^{1/d}$ in $d$-dimension. This led to the derivation of a 
super-Arrhenius scaling for the relaxation time of the form $\trel\asymp
\exp(\b^2/d\log 2)$. In a companion paper, using
mainly renormalization group ideas and electrical networks methods, we 
rigorously establish that instead $\trel\asymp \exp(\b^2/2d\log 2)$, a
result showing that the relaxation process cannot be 
quasi-one-dimensional. The above
scaling 
sharply confirms previous MCAMC (Monte Carlo with
Absorbing Markov Chains) simulations. Next we compute the relaxation
time at finite and mesoscopic length scales, and show a dramatic dependence on
the boundary conditions, yet another indication of key dimensional effects. Our final result is related to the
out-of-equilibrium dynamics. Starting with a single facilitating site at
the origin  we show that, up to length
scales $L=O(L_c)$, its influence propagates much faster (on a
logarithmic scale) along
the diagonal direction than along the axes directions. Such unexpected
result is due to a rather delicate balance between dynamical energy barriers and
entropic effects in the constrained dynamics.
}
\begin{document}
\maketitle
\section{Introduction and main results}
Kinetically constrained spins models  \cite{FH,JACKLE,Ritort} 
are stochastic 
particle/spin models, usually  defined in terms of a non-interacting
Hamiltonian, whose dynamics is determined by local rules encoding
a kinetic constraint. 
The main interest for these models (see e.g. 
\cite{Berthier,Ritort,SE2,GarrahanSollichToninelli,SE1,Cancrini:2006uu,Garrahan2003}) 
stems from the
fact  that KCMs, in spite of their simplicity, display many key dynamical
features of glass forming supercooled liquids: rapidly diverging
relaxation times as the temperature drops, super-Arrehenius behavior, dynamic heterogeneity
({\it i.e.} non-trivial spatio-temporal fluctuations of the
local relaxation to equilibrium) and aging, just to mention a
few.

One of the simplest models, showing all the above features and yet
being tractable even at a rigorous level, is the East model introduced
in \cite{JACKLE} and further analysed in
\cite{SE1,SE2,CDG,CMRT,FMRT-cmp,CFM,CFM-JSTAT,Blondel,East-cutoff,East-Rassegna}.
It is defined on a one dimensional integer lattice of $L$ sites, each
of which can be in state (or spin) $0$ or $1$, corresponding to
\emph{empty} or \emph{occupied} respectively. The spin configuration
$\eta$ evolves under Glauber type dynamics in the presence of the
kinetic constraint which forbids flips of those spins whose left
neighbor has spin one.  Each vertex $x$ waits an independent mean one
exponential time and then, provided that the current configuration
$\eta$ satisfies the constraint $\eta_{x-1}=0$, the value of $\eta_x$
is refreshed and set equal to $1$ with probability $1-c$ and to $0$
with probability $c$. The leftmost vertex $x = 1$ is unconstrained,
equivalently we could think of a fixed $0$ at the boundary $x=0$.
Since the constraint at site $x$ does not depend on the spin at $x$,
detailed balance is satisfied with respect to the product probability
measure $\pi:=\prod_{x}\p_x$, where $\pi_x$ is the Bernoulli
probability measure with density $c$ for the facilitating sites (the
vacancies). Taking $c$ equal to $e^{-\b}/(1+e^{-\b})$, where $\b$ is
the inverse temperature, then $\pi$ is the Gibbs distribution
associated to the non--interacting Hamiltonian given by the total
number of vacancies. In particular, small values of $c$ correspond to
low temperatures.

Recently the problem of dynamic heterogeneities and time-scales
separation in the East model  was rigorously
analyzed in \cite{CFM,CFM-JSTAT}\footnote{We warn the reader that in \cite{CFM,CFM2,CFM-JSTAT}
  the density $c$ of the facilitating sites is denoted by $q$.}, where the picture obtained in previous non rigorous work was corrected and extended. 
Dynamical heterogeneity is strongly associated to a \emph{broad spectrum}
of relaxation time scales which emerges as the result of a
subtle \emph{energy-entropy} competition. Isolated vacancies
with a string of $N$ particles to their left, cannot 
be updated unless the system injects enough
additional vacancies in a cooperative way in order to
unblock the target one. Finding the correct time scale on which this
unblocking process occurs requires a highly non-trivial analysis to
correctly measure the energy contribution (how many extra vacancies are
needed) and the entropic one (in how many ways the
unblocking process may occur). The final outcome is a very non trivial
dependence of the corresponding characteristic time scale on $c$ and
$N$. In particular it can be shown that the correct asymptotic of the relaxation time
$\trel^{\rm East}(L;c)$ as
$c\searrow 0$ takes the form ($n=\lceil \log_2L \rceil$ and we also allow
$L=+\infty$)
\begin{equation}
  \label{eq:1}
\trel^{\rm East}(L;c)\asymp
 \begin{cases} e^{\beta n } \times  n! \, 2^{-{n\choose 2}}, &  \text{$L \leq 1/c\,,$}
\\
e^{\beta^2 / 2 \log 2 }&  \text{$L \geq 1/c\,.$}
\end{cases}
\end{equation}
In the above formula $1/c$ is the mean inter--vacancy  distance and $n$ represents the minimal energy barrier between the ground state and the set of configurations with a vacancy at $L$. 
The above result came out as a surprise since the relaxation time is
determined not only by the ``standard'' energy barrier
contribution $1/c^n$ (cf. \cite{SE1,SE2,CDG}), but rather by the interplay between
energy and entropy (the latter being encoded in the new term 
$n! \,2^{-{n\choose 2}}$), a fact that was overlooked in previous work.

In several interesting contributions \cite{Gar1, Gar2,
  Garrahan2003,Keys2013, Gar3}
a natural generalization of the East dynamics
to higher dimensions $d>1$, in the sequel referred to as the \emph{East-like}
process, appears to play a key role in \emph{realistic}
models of glass formers. In $d=2$ the East-like process evolves 
similarly  to the East process but now the kinetic
constraint requires that the South \emph{or} West neighbor of the
updating vertex contains \emph{at least} one vacancy. In general
$\eta_x$ can flip if $\eta_{x-\vec e}=0$ for some $\vec e$ in the
canonical basis of $\bbZ^d$. 

\subsection{Infinite volume relaxation time}
A simple comparison with the East model shows that for any value of the vacancy density
$c$ the relaxation time of the East-like model on $\bbZ^d$, in the
sequel $\trel(\bbZ^d;c)$,  is finite. For example, in two dimensions the East-like model is in fact \emph{less}
constrained than a infinite array of East models, one for
each line parallel to the first coordinate axis, and this immediately gives rise to the conclusion. 
More subtle
is the problem of computing the asymptotic of $\trel(\bbZ^d;c)$ at low
temperature, a key intermediate step towards a qualitative and quantitative
description of dynamic heterogeneities in dimension greater than one. 

In \cite{Garrahan2003} it was assumed that the relaxation process of
the low temperature East-like models is \emph{quasi-one-dimensional},
{\it i.e.} it is determined by that of the East model on the equilibrium
scale\footnote{As equilibrium scale one can take the typical inter-vacancy distance under $\pi$.} in $d$-dimension $L_c= (1/c)^{1/d}$. In particular, it was argued
that the relaxation time of the East-like model scales like $\trel^{\rm East}(L_c;c)$. If one neglects  the
  entropic contribution in \eqref{eq:1}, as it was done in
  \cite{Garrahan2003}, the above
  assumption leads to a super-Arrhenius law of the form (recall that $\beta \simeq \log (1/c)$)
\begin{equation}
  \label{eq:2}
\trel(\bbZ^d;c)\asymp e^{\beta^2/d\log 2}.
\end{equation}
 Using the correct form $\eqref{eq:1}$ for $\trel^{\rm East}(L_c;c)$
gives instead
\begin{equation*}
\trel(\bbZ^d;c)\asymp e^{\beta^2(\frac{2d-1}{2d^2\log 2})}.
\end{equation*}
It turns out that both results are wrong, because of important
dimensional effects in the relaxation process of East-like models. Our first main result in 
 \cite{CFM2} shows in fact that:
\begin{maintheorem}
\label{th:main1}
As $c\searrow 0$  
\begin{equation}
\label{eq:4}
\trel(\bbZ^d;c) =e^{ \frac{\beta^2}{2d\log 2}(1+o(1))}
\end{equation}
and the $o(1)$
correction\footnote{Recall that
$f=O(g)$, $f=o(1)$ and $f=\O(g)$  mean that $|f|\le c |g|$ for
some constant $c$, $f\to 0$ and $\limsup |f|/|g| >0$ respectively.} is
$\O\left(\b^{-1}\log\beta\right)$ and $O(\b^{-1/2})$.
The above result can also be read as 
$$
\trel(\bbZ^d;c)= \trel^{\rm East}(\bbZ;c)^{\frac 1d (1+o(1))}.
$$ 
\end{maintheorem}
We discuss some interesting aspects of the derivation  of the above result later.
Notice that if we write the coefficient of
$\beta^2$ as $b/d$, then $b=1/(2\log 2)\approx 0.721$, a result that confirms
the value $b\approx 0.8$ found in 
simulations \cite[Fig.3]{Garrahan}  based on the ``Monte Carlo with Absorbing Markov Chains''
method \cite{Novot}.  
In finite or infinite  subsets of the lattice
$\bbZ^d$  the dimensional effects behind 
\eqref{eq:4}  depend strongly on the boundary conditions as explained below.
 \subsection{Finite volume relaxation time}
Consider  a finite box  with $L$ vertices on each edge in the positive
quadrant $\bbZ^d_+=\{(x_1,\dots,x_d)\in \bbZ^d:\ x_i\ge 0\}$ and containing the origin. 
 In order to ensure ergodicity of the
dynamics at least the spin at the origin must be unconstrained,
{\it i.e.} it should be free to flip. If this is the only unconstrained spin we say that we have
\emph{minimal} boundary conditions and we denote the associated relaxation time
 by $\trel^{\rm  min}(L;c)$. \emph{Maximal} boundary
conditions correspond instead to the case in which all the spins
in the box which belong to the coordinate hyperplanes $x_i=0$ are
unconstrained. In this case we write $\trel^{\rm max}(L;c)$ for the
associated relaxation time. Our second main result in \cite{CFM2} pins
down the asymptotics of the above relaxation times:
\begin{maintheorem}\ 
\label{th:main2}
Take $L$ depending on $c$ such that $\lim_{c\searrow 0}L=+\infty$. Then, as $c\searrow 0$, 
\begin{align}
  \label{eq:2}
\trel^{\rm max}(L;c)&=
\begin{cases}
e^{\left(n\beta - d{n\choose 2}\log 2\right)(1+o(1))}, &\text{$n\le \frac{\beta}{d\log 2}$},\\
  e^{ \frac{\beta^2}{2d\log 2}(1+o(1))}& \text{otherwise}.
\end{cases}\\
\label{eq:3}
\trel^{\rm min}(L;c)&=
\begin{cases}
e^{n\beta -{n\choose 2}\log 2  +n\log n +O(\beta)},&\text{$n\le \frac{\beta}{\log 2}$},\\
e^{ \frac{\beta^2}{2\log 2}+\frac{\beta \log \beta}{\log 2} 
  +O(\beta)}&\text{otherwise.}
\end{cases}
\end{align}
For $L$ fixed independent of $c$ the energy barrier dominates and the entropic reduction does not contribute to the leading order. 
  In particular for $L$ independent of $c$  as $c\searrow 0$,
\begin{equation*}
  \trel^{\rm min}(L;c)= e^{n\beta +O_n(1)},
\end{equation*}
where\footnote{The notation $O_n(1)$ means that the constant may depend on $n$. } $n = \lceil \log_2 \left( d(L-1) + 1\right)\rceil$.
 \end{maintheorem}
Notice that with minimal boundary conditions the relaxation time
scales exactly as the relaxation time of the East model given in
\eqref{eq:1}. The slowest mode in this case occurs along the coordinate axes.
\subsection{Persistence times} 
We conclude with a last result which highlights
once again some non-trivial dimensional effects in the
\emph{out-of-equilibrium} East-like dynamics. Consider the model in the positive
quadrant $\bbZ^d_+$ with minimal boundary condition (only the spin at the origin
is unconstrained) and starting from the 
configuration with no
vacancies. 
Let $T(x;c)$ be the \emph{persistence} time of $x\in \bbZ^d_+$, namely
the mean time it takes to create a vacancy
at $x$. 
The knowledge of the collection $T(x;c)$ as $x$ varies in $\bbZ_+^d$ gives some
insight on how  a wave of vacancies originating from a single one 
spreads out in space-time. Their analysis is therefore a key
step in order to understand the more complex phenomenon of time scale separation
and dynamic heterogeneities in a high-dimensional
setting.
In the last theorem we compute the low temperature scaling of $T(x;c)$ for sites $x$ which either belong to the diagonal of $\bbZ^d_+$ or to one of the
coordinate axes. Although the original result in \cite[Theorem 3]{CFM2} covers
quite precisely \emph{all} scales up to the equilibrium scale $L_c$, in order to
outline our main finding we describe it only for $|x|\approx L_c$, $L_c=(1/c)^{1/d}$,
and $|x|=O(1)$ and we only give the leading term in $\b$ in the
asymptotics without specifying the error.
\begin{maintheorem}\ 
\label{th:main3}
Let $v_*=(d L_c,0,\dots,0)$ and $v^*=(L_c,L_c,\dots,L_c)$ so
  that $v_*$ belongs to the first coordinate axis and $v^*$ to the diagonal.  
Then, as $c\searrow 0$, the persistence time satisfies
    \begin{align}
      \label{hitmin}
  T(v_*;c)&\asymp \trel^{\rm East}(L_c;c)\asymp \trel^{\rm
    East}(\bbZ;c)^{(2d-1)/d^2},\\
      \label{hitmax} 
T(v^*;c)&\asymp \trel(\bbZ^d;c)\asymp \trel^{\rm East}(\bbZ;c)^{1/d}.
    \end{align} 
Fix $n\in \bbN$ and let $x\in \bbZ_+^d$ be such that\footnote{The $\ell_1$-norm of
  $x=(x_1,\dots,x_d)$ is given by $\|x\|_1:=\sum_{i=1}^d|x_i|$. }
  $\|x\|_1+1\in [2^{n-1},2^n)$. Then,
\begin{align}
  \label{hitmin2}
  T(x;c)= e^{n\beta +O_n(1)}.
\end{align}
\end{maintheorem}
\begin{figure}[htbp!]
\centering
\includegraphics[width=.58\linewidth]{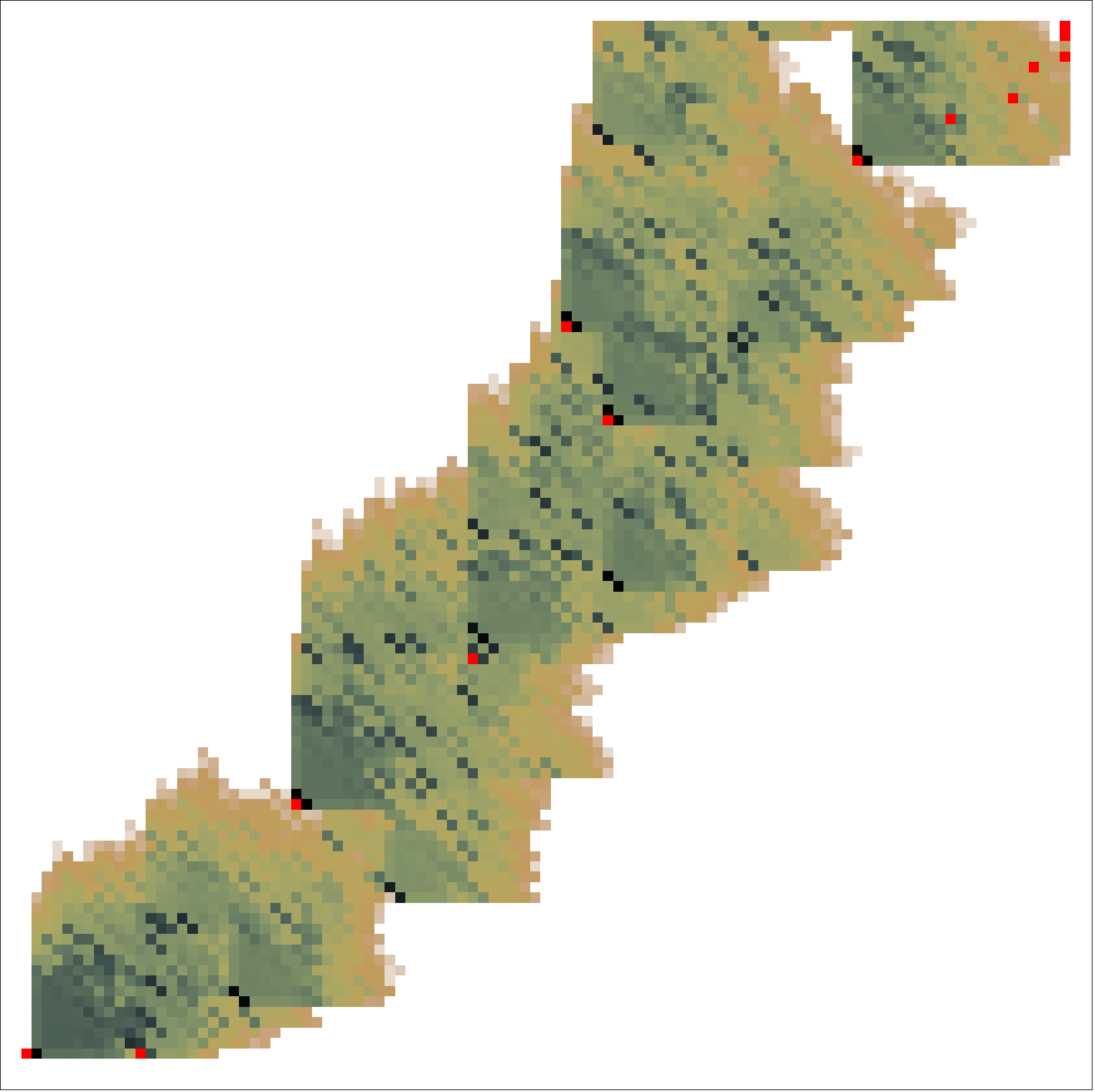} 
\includegraphics[width=.58\linewidth]{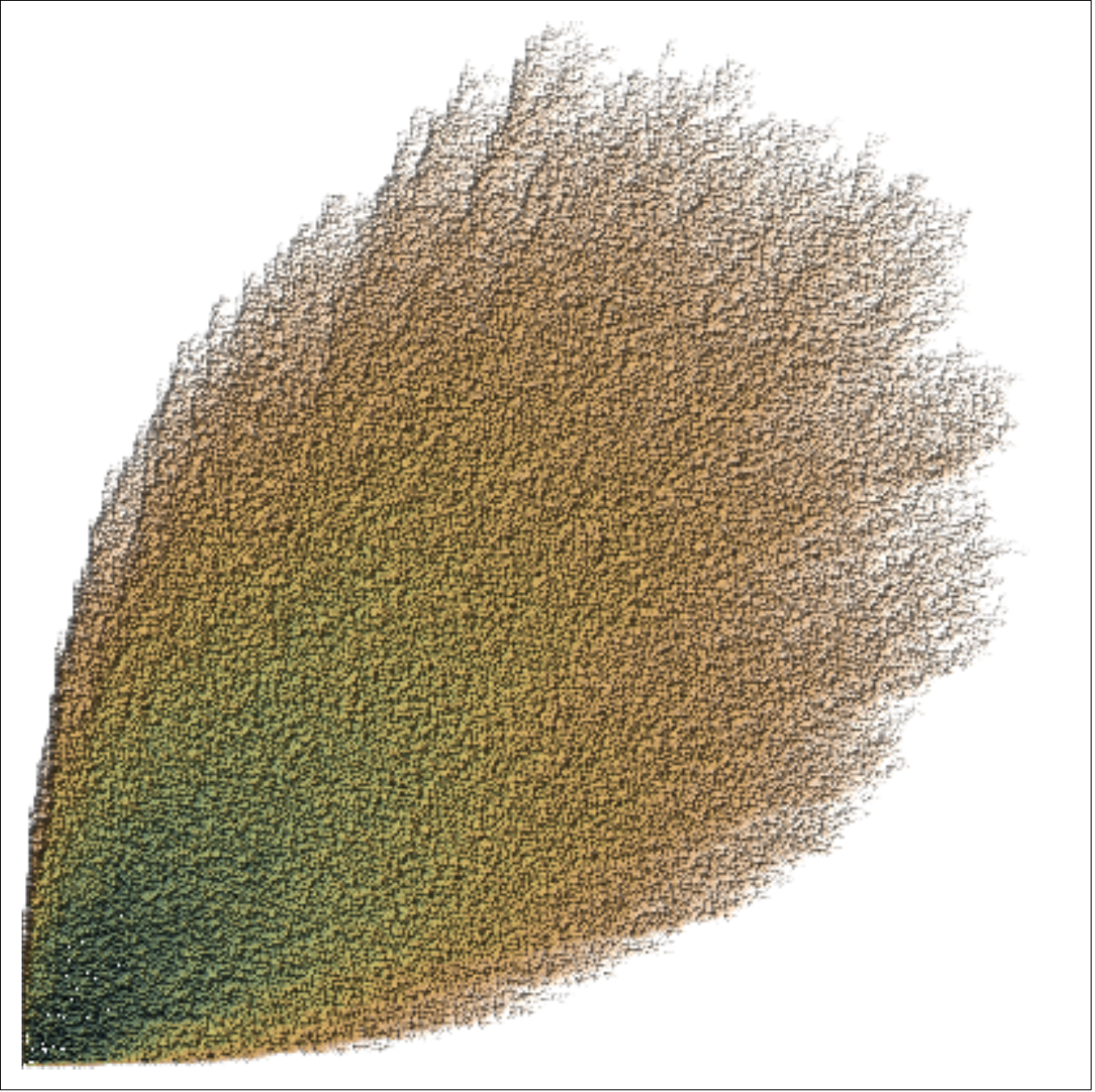} 
\caption{A snapshot of the East-like process with minimal boundary conditions
    and initial condition with no vacancies. White vertices  have
    never been updated, the darker the color the more times the site
    has refreshed.  (TOP) $c=0.002$, snapshot taken at the
    first time a vacancy is present at the top-right corner $(100,100)$ ($t \simeq 2.5\times 10^{12}$). Red dots denote the vacancies present in the
    snapshot.  (BOTTOM) $c=0.25$, $L=400$ and $t=9 \times 10 ^ {3}$. }
\label{fig:1bis}
\end{figure}
 We conclude this first part with two observations. Notice
 that we (intentionally) always compare the persistence time of
vertices with equal $\ell_1$-norm. If in fact the
relaxation process for the East-like model was quasi-one dimensional,
then the asymptotics of $T(x;c)$ should be determined by 
the minimal
length of a path connecting the origin to $x$, at least at the
logarithmic level. When the distance of $x$ from
the initially unconstrained spin at the origin is $O(1)$, that is
indeed correct as shown by \eqref{hitmin2}.
Instead, when the distance of $x$ from the origin is large (diverging
as $c\searrow 0$),
the $\ell_1$-norm is no longer sufficient and the direction matters
dramatically. One could wonder whether another norm would be more
relevant at this scale. Indeed $v^*$
has much smaller Euclidean norm than $v_*$ ($\sqrt{d}L_c$ compared to
$dL_c$) and that might be responsible for the
difference in the corresponding persistence times. However, the full result of
\cite[Theorem 3]{CFM2} shows that this is not the case and that, for any $\l,\l' >0$ independent of
$c$, $T(\l v_*;c)$ is (logarithmically) much shorter
than $T(\l' v^*;c)$ as $c\searrow 0$. Of course, if $\l,\l'$
\emph{depend} on $c$ the situation may change completely. Theorem 3 in
\cite{CFM2} indeed implies for example that  $T(v^*;c)\asymp T(c^\a\, v_*;c)$ if $\a=
1/d+\sqrt{1-1/d}-1$.
  Lastly, in case (1) we observe that the persistence time of the
vertex $v^*$ is much shorter (on a logarithmic
scale) then the relaxation time of a finite
box of side $dL_c$ and minimal boundary conditions (cf. Theorem \eqref{th:main2}). In this case, contrary to
what happens for the East model, the system is able to bring a vacancy
very far from the initial one \emph{without} reaching
equilibrium in the corresponding
box (see Figure \ref{fig:1bis}).
\section{Outline of the methods}
We use several techniques to derive the above results. As a
preliminary step we
exploit the monotonicity in the constraints of the quadratic form associated to
the generator of the master equation to derive the inequalities
\begin{align}
\trel(\bbZ^d;c)&\le \trel^{\rm East}(\bbZ;c)   \label{eq:5bis},\\
\trel^{\rm
  max}(L;c)\le \trel^{\rm min}(L;c),&\;
\trel^{\rm East}(L;c)\le \trel^{\rm min}(L;c)\label{esami},
\end{align}
and the fact that  $\trel^{\rm max}(L;c),\, \trel^{\rm min}(L;c)$ are
both increasing in $L$. It is worth noticing that \eqref{esami}
together with \eqref{eq:1} leads to the correct lower bound asymptotics in \eqref{eq:3}. Next we use \cite[Prop. 2.13]{CMRT} together
with the monotonicity in $L$ of $\trel^{\rm max}(L;c)$ to get
that\footnote{Notice that the infinite volume relaxation time is \emph{not}
  given by the large system limit with minimal boundary conditions
  (cf. \eqref{eq:4} and \eqref{eq:3}).}
    $\trel(\bbZ^d;q)=\lim_{L\to \infty}\trel^{\rm
      max}(L;q)$. Unfortunately no
    monotonicity argument is available for the persistence times since
    they are not characterized uniquely by a variational principle involving the
    quadratic form of the generator. 
The proof of our results then rely on four main ingredients:\\
- comparison with the East model on a spanning tree to \emph{upper
  bound} $\trel^{\rm min}(L;c)$;\\
- a renormalization group
    approach, involving a sequence of coarse-grained auxiliary dynamics, to 
    \emph{upper bound} $\trel(\bbZ^d;c)$ and
    $\trel^{\rm max}(L;c)$;\\
- an algorithmic construction of an
    efficient bottleneck in the configuration space to \emph{lower
      bound} $\trel(\bbZ^d;c)$, 
    $\trel^{\rm max}(L;c)$ and the persistence time $T(x;c)$;\\
- resistors network techniques to \emph{upper bound} the persistence times.
\subsection{East model on a spanning tree}
Choose a rooted, oriented spanning tree $\cT$ for the box $\L=[0,L-1]^d$ with
root at the origin and edges oriented as the canonical basis of
$\bbZ^d$. On $\cT$ we consider a constrained East dynamics in which
the spin at $x$ is free to flip if either $x$ is the root or there is a
vacancy at the $\cT$-ancestor of $x$. This new dynamics is more
constrained than the East-like dynamics in $\L$ with minimal boundary
conditions. Hence $\trel^{\rm min}(L;c)\le \trel(\cT;c)$, where
$\trel(\cT;c)$ denotes the relaxation time of the $\cT$-chain. In turn, as shown in
\cite[Th. 6.1 and eq. (6.3)]{Praga}, $\trel^{\rm
  min}(\cT;c)$ is smaller than the relaxation time of East process on
the longest branch of $\cT$. To upper bound the latter it is enough to
apply \eqref{eq:1}. 
\subsection{Block dynamics and renormalization group}
Given $\ell=2^n$ and $x \in \bbZ^d$ let $\L(x, \ell):=\ell x + [0, \ell-1]^d$. We
partition the lattice  $\bbZ^d$ into (disjoint) blocks $\L( x, \ell)$ and
introduce an auxiliary constrained \emph{block dynamics}
on $\{0,1\}^{\bbZ^d}$ which mimics the East-like dynamics on a
coarse-grained scale. In each block $\L(x,\ell)$ with rate one the spin
configuration inside the block is refreshed to a new one sampled from
the equilibrium distribution $\pi$  provided that at least one
neighboring block $\L(x-\vec e, \ell)$, $\vec e$ being a vector in the canonical basis  of
$\bbZ^d$, contains a vacancy. 
 Note that the block dynamics reduces to the East--like process when
 $\ell=1$. Writing $c_*\simeq e^{-\beta_*}$  for  the probability $1- (1-c)^{\ell^d} $
 that the block constraint is satisfied at $x$,  it is simple to
 verify that the relaxation time  $\trel^{\rm block} (\bbZ^d;c)$ of
 the block dynamics coincides with $\trel (\bbZ^d; c_*)$. The idea at
 this point is to search for an explicit rescaling function
 $F(\ell;c)$ such that
$$
\trel(\bbZ^d;c)\simeq F(\ell,c) \trel^{\rm block} (\bbZ^d;c)=F(\ell,c) \trel(\bbZ^d;c_*).
$$
Once such an equation is available then, by choosing appropriately the
block side $\ell$ as a function of the density $c$, one
could hope to solve for $\trel(\bbZ^d;c)$. 

Clearly the sought function $F(\ell,c)$ should (roughly) be the time scale
over which a single vacancy in one of the blocks $\L(x-\vec e,\ell)$,
whose presence is guaranteed by the block constraint,
is able to induce equilibrium inside the block $\L(x,\ell)$.  
Although we could not implement exactly the above program, we
rigorously show in \cite{CFM2} that 
\begin{equation}\label{normale2}
\trel (\bbZ^d; c)\leq \k   \trel^{\rm min} (3 \ell; c)\,\trel (\bbZ^d;c_*)
\end{equation}
for some constant $\k$ depending only on $d$.
Using \eqref{eq:3} of Theorem \ref{th:main2} we know that  
\begin{equation}\label{spine1}
\trel^{\rm min}(3\ell;c) 	\leq
e^{(n\beta -\frac{ \log 2}{2} n^2)(1+o(1))}
   \end{equation}
for any $\ell \leq L_c$ large enough. 
Suppose now that 
\begin{equation}\label{spine3}
\trel (\bbZ^d;c) \leq e^{\l \frac{\beta^2}{2 \log 2 }(1+o(1))}
 \end{equation}
for some $\lambda \in
(1/d, 1]$. Using \eqref{eq:5bis} and the exact asymptotics of
$\trel^{\rm East}(\bbZ;c)$ (cf. \eqref{eq:1}) we indeed know that \eqref{spine3} holds for $\l=1$.
Then we plug \eqref{spine1} and \eqref{spine3}\footnote{With $c$ replaced
by $c_*$ in \eqref{spine3} and therefore with $\b$ replaced by
$\b_*\simeq \b - d n\log 2$ for $n$ large and $c$ small.} into the r.h.s. of \eqref{normale2} and
optimize over the free scale $\ell \leq L_c$.  The conclusion is that
\eqref{spine3} is actually valid for a smaller constant 
$$
\lambda'=
\frac{2d\l -1-\l}{d^2\l-1} \in  (1/d, 1].
$$
It is easy to verify that the  map $\lambda \mapsto \lambda'$, $\l\in [1/d,1]$, has an
attractive quadratic fixed point at $\l_c=1/d$. Thus, starting from
$\l=1$ and by iterating a large number of times the above recursion we get \eqref{spine3}
with $\l=1/d$.

\begin{figure}[htbp!]
\centering
\begin{overpic}[scale=0.25]{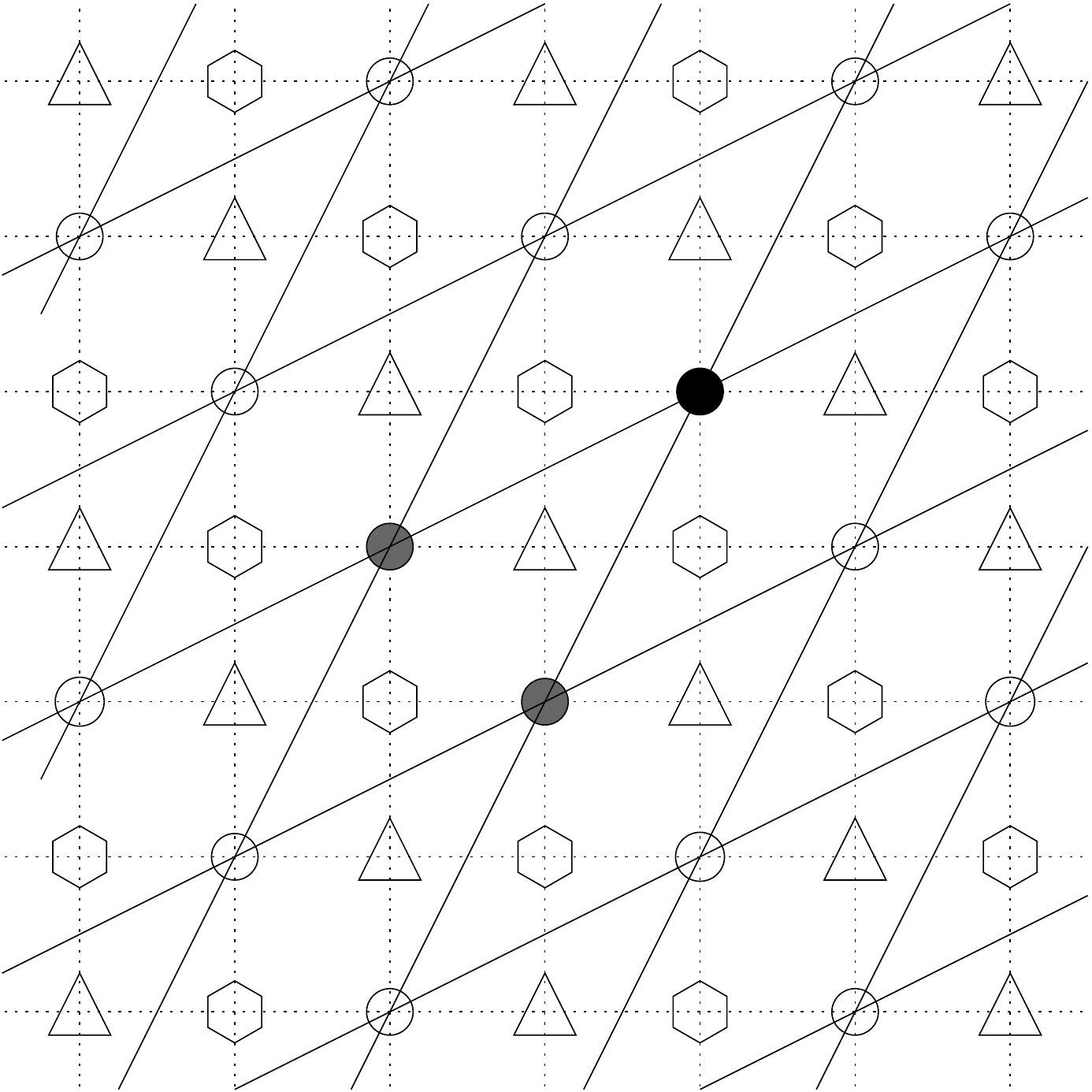}
\put(32,55) {$y$}
\put(58,68) {$x$}
\put(46,40) {$z$}
\end{overpic}
\caption{Partition of $\bbZ^2$ in three sub--lattices distinguished by
  different geometric shapes. For
  circles, also the edges of the sub-lattice have been drawn. In the
  Knight Chain, the block $\L(x,\ell)$ associated to $x$ only looks
  at the blocks $\L(y,\ell),\, \L(z,\ell)$, with $y,z$ in the same
  sub-lattice of $x$.}
\label{DDD}
\end{figure}

We point out that in the rigorous derivation of \eqref{normale2} we
were forced to use a slightly different block dynamics. 
Indeed,  the presence  of  
 vacancies only inside some box $\L(x-\vec e,\ell)$ is, in general, not
 enough to guarantee equilibration of the East--like process inside
 the block $\L(x,\ell)$. Therefore, in some sense the
 East--like process and the above block dynamics are not directly
 comparable. 
To overcome this problem the
 block constraint for $\L(x,\ell)$ was modified to 
require the presence of some
 vacancy in (at least) one block of the form $\L(x-(1,1, \dots, 1)-\vec e,
 \ell)$. In e.g. $d=2$ we thus ask for some vacancy in $\L(x-(2,1),\ell)$ or
in  $\L(x-(1,2), \ell)$.  In analogy with the chess piece,  we call  the resulting 
dynamics the \emph{Knight Chain}. It is easy to check that the Knight Chain
consists of $d+1$ independent chains, each one isomorphic to the
previously defined block dynamics (see Figure \ref{DDD}). 
In particular, the infinite volume  relaxation time of the Knight
Chain coincides with $\trel^{\rm block}(\bbZ^d;c)=\trel(\bbZ^d;c_*)$.

\subsection{Bottleneck inequality} To get a lower bound on $\trel^{\rm
max}(L;c)$ we 
use the Rayleigh-Ritz variational principle for the spectral gap of
the generator of the master equation. Restricting to test functions
of the form $f(\eta)=1$ if $\eta \in A$ and $f(\eta)=0$ otherwise, for
some subset $A$ in the spin configuration space, one easily gets 
  \begin{equation}\label{ciliegia}
T^{\rm max} _{\rm rel}(L)  \geq  L^{-d} \frac{ \pi(A) \pi(A^c) }{\pi( \partial A) }\,,
\end{equation}
where the boundary $\partial A$ is given by the configurations $\h $
in $A$ such that the East-like process can jump from $\eta$ to $A^c$.
When $\pi(A)\le 1/2$ and $\pi (\partial A)\ll \pi(A)$, one says that
$\partial A$ is a bottleneck for the dynamics. The problem is then to
find a set $A$ leading to the correct asymptotics in \eqref{eq:2}.

 We provide an  algorithmic construction of such a set $A$. Given a configuration $\h$
 on $[1,L]^d$ with a vacancy at $x=(x_1, x_2, \dots, x_d)$, we  define
 the gap of this vacancy as the minimal $\ell_1$-distance between $x$ and
 $y \in  \prod _{i=1}^d [1,x_i]$ such that  there is  a vacancy at
 $y$ (in $\eta$ or  in the maximal boundary conditions).
The algorithm proceeds as follows.
Starting from a spin configuration $\eta$ remove from $\eta$ all vacancies with
gap equal to $1$, then remove from the resulting configuration all
vacancies with gap equal to $2$, and so on until all vacancies with
gap equal to  $L-1$ have been removed. At this point the algorithm stops. 
Due to the maximal boundary conditions there are only two possible
final spin configurations: the fully occupied 
one and $\mathbf{1}0$, the configuration with a single vacancy at
$(L,L, \dots ,L)$. 
Our set $A$ is then given by those spin configurations $\eta$ for which the
algorithm outputs $\mathbf{1}0$. The estimate of the ratio $\pi(A) \pi(A^c) /\pi( \partial A) $ is not trivial and is based on the fact that  configurations in $\partial A$ must have at least $\lceil \log_2 L \rceil$ vacancies whose locations have  special geometric properties. 
\subsection{Resistor network} 
We use a standard analogy between electrical networks and
Markov processes satisfying  the detailed balance condition \cite{DS}.
In particular, we map the East--like process on the box $[1,L]^d$,
with minimal boundary conditions,  to a resistor network whose nodes
are given by the configurations $\eta$ on $[1,L]^d$.  Calling
$\cK(\h,\h')$  the probability rate for a jump from  $\h$ to $\h'$, to each unordered pair $\h,\h'$ with $\cK(\h,\h')>0$  we attach a resistor with conductance 
 $\cC(\h,\h')$ given by 
\begin{equation}
  \label{eq:conduct}
  \cC(\h,\h') = \pi(\h)\cK(\h,\h') = \pi(\h')\cK(\h',\h)\,.
\end{equation}

Given $x \in [1,L]^d$ we set $B_x:=\{\eta\,:\, \eta_x =0\}$ and we let $\mathbf{1}$  denote the configuration with no vacancy. We write $R(x) $ for the effective resistance between 
$\mathbf{1}$ and $B_x$, given by the inverse current intensity when putting  potential one on    $\mathbf{1}$ and zero on $B_x$.  Then the  equilibrium potential at $\eta$ equals the probability 
$\bbP_\eta( \t_{\mathbf{1} }< \t_{B_x})$
for the East-like  process starting from $\eta$ to hit $\mathbf{1}$ before $B_x$ and the persistence 
time $T(x;c)$ satisfies
\begin{equation}\label{stanchi} T(x;c)= R(x) \sum _{ \eta \not \in B_x} \pi (\eta) \bbP_\eta( \t_{\mathbf{1} }< \t_{B_x})\,.
\end{equation}
The above identity implies that $T(x;c) \asymp R(x)$ if $L \leq L_c$.
To bound  $R(x)$ above we use Thomson's principle which states  that  $R(x)$ is the minimal energy $\cE( \theta)= \frac{1}{2}\sum_{ \h,\h'} \theta(\h, \h')^2/ \cC(\h,\h')$   over unit flows $\theta$ from $\mathbf{1}$ to $B_x$. A  unit flow $\theta$ is an  antisymmetric function on the set of ordered pairs $(\h,\h')$ with $\cK(\h,\h')>0$, which is divergence free on $ ( \mathbf{1} \cup B_x )^c$ and such that the flow exiting $\mathbf{1}$ is unitary.  Thomson's principle reduces the problem to finding a unit flow whose energy has the expected asymptotic.

To construct such a flow we use  a hierarchical 
procedure which, for $L=O(1)$ and $d=1$,  resembles the hierarchical method used to estimate  the energy barrier that must  be overcome  in order to bring a vacancy at distance $L$ \cite{SE1,CDG}.

First we introduce a unit flow $\theta_{y}$ from $\mathbf{1}$ to $B_x$ (passing through a  point $y$  around $x/2$) in three steps.
 Consider  the equilibrium unit flow $\phi_y$ from $\mathbf{1}$ to $B_y$  
which has energy $R (y)$.  Consider then the reversed flow $\hat \phi_y$ keeping the vacancy at $y$ fixed and removing all other vacancies, defined as 
$\hat \phi _y( \eta, \eta')=  \phi_y( \eta'^y, \eta^y)$  if $\h,\h' \in B_y$ and zero otherwise ($\s^y$ is obtained from $\s$ by a single spin--flip at $y$). Note that $\phi_y + \hat \phi_y$ is a unit flow from $\mathbf{1}$ to the configuration $\mathbf{1}0_y$ with a single vacancy at $y$. Finally, let $\tilde \phi_y$ be the unit flow from $\mathbf{1}0_y$ to $B_x$  which mimics on $\D=[y+1,x]\times [y,x]^{d-1}$ the equilibrium unit flow from the filled configuration on $\D$ (no vacancy) to the set of configurations on $\D$ with a vacancy at $x$.
Finally we define the flow $\theta_y= \phi_y+ \hat \phi_y + \tilde \phi_y$, which is indeed a unit flow from $\mathbf{1}$ to $B_x$. Since $ \phi_y, \hat \phi_y , \tilde \phi_y$ have disjoint supports, the energy of $\theta_y$ is the sum of the individual energy, which is related to $R(y)$. At the end, at least for $y \approx x/2$, we get
\begin{equation}
\label{Paul:1}
R(x)\leq \cE( \theta_y)  \leq R(y) + \frac{2}{ c} R(y)
\end{equation}
(the factor $1/c$ is due to  the vacancy at $y$).

 The above recursive inequality \eqref{Paul:1} is  indeed not sufficient to get the correct asymptotic for the upper bound of $R(x)$, since it disregards relevant entropic effects. In fact, to create a vacancy at $x$, 
  the system
can firstly bring a zero to any point close to the midpoint of $x$,
potentially at the cost of a small number of extra vacancies, which gives rise to many more paths.
We therefore replace $\theta_y$ with a local average $ N^{-1} \sum_{y\in V_x} \theta_y$, where $V_x$ is a box around $x/2$ with $N$ points. Then, instead of \eqref{Paul:1}, we get
 (see \cite[Lemma
7.1]{CFM2})
\begin{equation}
  \label{Paul:2}
R(x) \leq \frac{C}{N} \sum_{y\in V_x} R(y) +  \frac{C}{c
  N^2}\sum_{y\in V_x} R(y) 
\end{equation}
for some universal constant $C$.
It then remains to find a good choice of the box $V_x$:  increasing the
size $N$  of the box $V_x$ accounts for more entropy, on
the other hand if $V_x$ is too large then some of the points $y$
become close to  the coordinate axes, thus leading to a very large resistance $R(y)$ since 
there is a smaller entropy ($R(y)$ would approach the very large  1D persistence time $T(y;c)$).
Applying the above bound recursively and choosing the box $V_x$
carefully we arrive at the upper bound on the hitting time stated in
Theorem \ref{th:main3}.
\acknowledgments
We would like to thank the partipants of the workshop ``Glassy
Systems  and Constrained Stochastic Dynamics'' (Math. Dept, Warwick Univ. 2014)
and in particular J.P. Garrahan for several stimulating discussions. 

\end{document}